# Modification of the quasi-lattice theory for liquid alloys on the basis of varying the coordination number and its application to Al-Sn, Al-Zn and Sn-Zn

**O.M. Oshakuade\*** and **O.E. Awe**

Department of Physics, University of Ibadan, Ibadan, Nigeria

## Abstract

The existing quasi-lattice theory for liquid alloys (QLT), which has been extensively used by many researchers, has been modified by incorporating the knowledge of composition and temperature-dependent coordination numbers. The modified QLT was then used to compute the enthalpy of mixing, the entropy of mixing, concentration fluctuations, Warren-Cowley short range order parameter, surface concentrations and surface tensions of liquid Al-Sn, Al-Zn and Sn-Zn systems, which are the binary sub-systems for Al-Sn-Zn. The effect of the approximation of coordination number in the existing QLT was also investigated and was found to be insignificant when coordination number is 10. This work has provided a more physically realistic quasi-lattice theory, and has contributed to the knowledge on the binary subsystems of Al-Sn-Zn and has also set a foundation for the application of quasi-lattice theory on Al-Sn-Zn and other ternary systems.

Corresponding author. Tel.: +2348038557420; E-mail: om.oshakuade@ui.edu.ng, gbengaoshaks@yahoo.com

# 1   Introduction

Quasi-lattice theory for compound forming liquid alloys (QLT) [1] can be described as an improvement over the Complex formation model (CFM) [2], because QLT provides a better statistical mechanical background to CFM. In addition, QLT reduces to the Conformal solution and Flory's approximations of CFM [2] when Coordination number ($Z$) has a value of two and infinity, respectively. Amongst several existing thermodynamic models, QLT is very relevant for its ability to describe alloy systems in terms of the relative bonding energies between atom pairs and for its success in the study of concentration-concentration fluctuations in the long wavelength limit ($Scc(0)$) of liquid alloys [1,3].

In many studies, coordination numbers of liquid alloys are taken as constant, regardless of their components' structures, composition and temperatures [4–8]. Whereas, coordination number of metals (and other elements) is known to depend on their crystal structures [9,10]. Experimental results have revealed that the coordination numbers of molten metals depend on their temperature [11,12]. Theorists [13,14] have related the coordination numbers of liquid metals to their pair distribution functions, which is temperature-dependent [15]. Considering the non-crystalline nature of alloys, Elliot [16] had suggested composition-dependent coordination numbers. When more physically realistic Coordination numbers ($Z$) are applied in the QLT [1], they are expected to give the QLT a more significant advantage over the CFM [2]. Thus, an improved estimate of $Z$ is expected to improve the quality of the results obtained using QLT.

Therefore, while the use of Z =10 for all liquid alloys (and at all temperatures) may be a convenient approximation [3], it can be modified to incorporate more recent knowledge [9,10,13,14]. Furthermore, the coordination numbers in liquid alloys, unlike previous work in QLT, are expected to depend on temperature [15], composition [16] and crystal structures of components [9,10] of the system.

In this work, we have attempted to modify the existing QLT [1] by incorporating the contributions of temperature, composition, and structures of system's components to the coordination number in the existing QLT. The resulting modified QLT is then used to study Al-Sn, Al-Zn and Sn-Zn systems. The interest in the aforementioned systems is motivated by the fact that they are binary subsystems of Al-Sn-Zn ternary system, a Pb-free solder candidate [17–20]. There are existing experimental thermodynamic activity data for Al-Sn, Al-Zn, Sn-Zn and Al-Sn-Zn systems at 973, 1000, 750 and 973 K, respectively [17,21]. Experimental thermodynamic activity data for Al-Zn and Sn-Zn at 973 K were deduced from EMF data [22,23] and subsequent integration of Gibbs-Duhem equation [24,25]. In this work, thermodynamic studies were carried out on the three binary systems at temperatures corresponding to their known binary and super-ternary (973 K) experimental information. The effects of the approximation of coordination numbers on the prediction of thermodynamic properties were also studied.

The paper is laid out as follows: the next section gives concise details on QLT, its modification, and application to enthalpy of formation, entropy of formation, concentration fluctuations, Warren-Cowley chemical short range order parameter and surface properties calculation, section 3 contains results and discussions on the computed thermodynamic properties, while section 4 contains summary and conclusions.

## 2 Theoretical framework

### 2.1 Average coordination number $(\widehat{Z})$

Tao [14] developed a model for the estimation of temperature-dependent coordination numbers for pure liquid metals ($Z_i$) using the physical properties of metals, as defined in Eq. (1).

$$Z_i = \frac{4\sqrt{2\pi}}{3} \left( \frac{r_{mi}^3 - r_{0i}^3}{r_{mi} - r_{0i}} \right) \frac{0.6022\, r_{mi}}{V_{mi}} \exp\left( \frac{\Delta H_{mi}(T_{mi} - T)}{Z_c R\, T\, T_{mi}} \right) \qquad (1)$$

Where $T$ is the temperature of the system under study, $\Delta H_{mi}$ and $T_{mi}$ are the melting enthalpy and melting temperature of component $i$, respectively, $Z_c$ is the close-packed coordination number of magnitude 12, $V_{mi}$ is the molar volume of component $i$ at $T$, $r_{0i}$ and $r_{mi}$ are the beginning and first peak values of radial distance distribution function near the melting point of liquid metal $i$, respectively. The relevant parameters used in this work to estimate $Z_i$ are presented in Table 1 [26].

Table 1: Calculated $Z_i$ of each component and essential parameters related to its calculation

| Metal i | $V_{mi}$ [a] (cm³/mol) | $T_{mi}$ [a] (K) | $\Delta H_{mi}$ [a] (kJ/mol) | $r_{mi}$ [a, b] | $r_{0i}$ [a, b] | $Z_i(T)$ [c] 750 K | 973 K | 1000 K |
|---|---|---|---|---|---|---|---|---|
| Al | 11.30 [1 + 0.00015 (T - 933)] | 933 | 10.46 | 2.78 | 2.28 | - | 9.4385 | 9.3734 |
| Sn | 17.00 [1 + 0.000087 (T - 505)] | 505 | 7.07 | 3.14 | 2.68 | 8.8510 | 8.4999 | - |
| Zn | 9.94 [1 + 0.00015 (T - 693)] | 693 | 7.28 | 2.66 | 2.16 | 9.2641 | 8.7690 | 8.7173 |

[a] These parameters were obtained from [26]
[b] Unit is $10^{-8}$ cm
[c] The various values for different temperatures were obtained using Eq. (1)

The Average coordination number, $\hat{Z}$, introduced in Eq. (2), was intended to account for the composition dependence of $Z$ in liquid alloys [16].

$$\hat{Z} = \sum_{i=1}^{2} Z_i c_i \qquad (2)$$

Where $Z_i$ is the coordination number of pure element $i$ (defined by Eq. (1)) and $c_i$ is the molar concentration (or mole fraction) of component $i$ in the system.

Expanding Eqs. (1) and (2) for a generic "$A - B$" binary system, gives Eq. (3), an expression for average coordination number that is dependent on crystal structure of components, composition and temperature.

$$\hat{Z}(c_A, T) = \frac{4 \times 0.6022 \sqrt{2\pi}}{3} \left[ \left( \frac{r_{mA}^3 - r_{0A}^3}{r_{mA} - r_{0A}} \right) \frac{r_{mA}}{V_{mA}} \exp\left( \frac{\Delta H_{mA}(T_{mA} - T)}{Z_c R T T_{mA}} \right) c_A \right. \\ \left. + \left( \frac{r_{mB}^3 - r_{0B}^3}{r_{mB} - r_{0B}} \right) \frac{r_{mB}}{V_{mB}} \exp\left( \frac{\Delta H_{mB}(T_{mB} - T)}{Z_c R T T_{mB}} \right) (1 - c_A) \right] \qquad (3)$$

For non-crystalline systems (such as alloys), the use of average coordination numbers, $\hat{Z}$, (instead of the nominal coordination numbers) will be more appropriate [16,27,28]. In this work, the Average coordination number $(\hat{Z})$ is adopted to study alloy systems.

## 2.2 Quasi-lattice theory for compound forming liquid alloys (QLT)

The QLT by Bhatia and Singh [1], like CFM, is built on the assumption that $N_A$ moles of element $A$ and $N_B$ moles of element $B$ will mix to form a pseudo-ternary mixture comprising of $n_1$ moles of unassociated atoms of element $A$, $n_2$ moles of unassociated atoms of element $B$ and $n_3$ moles of $A_\mu B_\nu$ complex. Where $\mu$ and $\nu$ are small integers obtained from stoichiometric information while $c_1$ and $c_2$ are the molar concentrations of elements $A$ and $B$, respectively. If the total number of moles in the mixture is $N$, and the atoms are conserved, Eqs. (4) - (8) will be valid. The average coordination number $(\hat{Z})$ described in Section 2.1 (Eq. (3)) is used in this section to replace the nominal $Z$ in previous works.

$$N_A + N_B = N \tag{4}$$

$$N_A = N\, c_1 \tag{5}$$

$$N_B = N - N_A = N - N\, c_1 = N(1 - c_1) \tag{6}$$

$$n_1 = N\, c_1 - \mu\, n_3 \tag{7}$$

$$n_2 = N\,(1 - c_1) - \nu\, n_3 \tag{8}$$

The expression for $G_M$ given as Eq. (9), can be written as Eq. (10) [2,25].

$$G_M = G - N\, c_1\, G_1^{(0)} - N\, c_2\, G_2^{(0)} \tag{9}$$

Where $G_i^{(0)}$ is the chemical potential for specie $i$ in the mixture, $G$ is the free energy of the mixture.

The $G_M$ can also be defined in terms of the pseudo-ternary constituents as Eq. (10) [2,25]

$$G_M = -n_3 g + G' \tag{10}$$

Where $g$ is the free energy of formation of chemical complex, the term "$-n_3 g$" denotes the lowering of free energy as a result of chemical complex formation, $G'$ is the free energy of mixing of the pseudo-ternary mixture assumed to interact weakly, and $G'$ depends on the underlying relevant theory being applied to study weakly interacting mixtures. Generally, $g$ and $G'$ is defined as Eqs. (11) and (12).

$$g = \mu\, G_1^{(0)} + \nu\, G_2^{(0)} - G_3^{(0)} \tag{11}$$

$$G' = G - \left(n_1\, G_1^{(0)} + n_2\, G_2^{(0)} + n_3\, G_3^{(0)}\right) \tag{12}$$

For an ideal mixture, $G'$ can be defined as Eq. (13), where $R$ and $T$ are the molar gas constant and temperature, respectively.

$$G' = RT\left(n_1 \ln\frac{n_1}{n_1+n_2+n_3} + n_2 \ln\frac{n_2}{n_1+n_2+n_3} + n_3 \ln\frac{n_3}{n_1+n_2+n_3}\right) \tag{13}$$

Assuming only one type of complex is formed, the Bhatia and Singh QLT [1] gives the expression for $G_M$ as Eq. (14).

$$\begin{aligned}G_M = -n_3 g + RT\bigg[&n_1 \ln\frac{n_1}{N} + n_2 \ln\frac{n_2}{N} + n_3 \ln\frac{(\mu+\nu)n_3}{N} \\ &- \frac{1}{2}\widehat{Z}n_3(\mu+\nu-\zeta)\ln\frac{\mu+\nu}{\mu+\nu-\zeta} - \frac{1}{2}\widehat{Z}\,\aleph \ln\frac{\aleph}{N}\bigg] \\ &+ \frac{1}{\aleph}(n_1 n_2 \nu_{12} + n_1 n_3 \nu_{13} + n_2 n_3 \nu_{23})\end{aligned} \tag{14}$$

Where $\zeta$ and $\aleph$ are defined in Eqs. (15) and (16), respectively.

$$\zeta = \frac{2(\mu+\nu-1)}{\widehat{Z}} \tag{15}$$

$$\aleph = n_1 + n_2 + \left(\mu+\nu - \frac{2(\mu+\nu-1)}{\widehat{Z}}\right)n_3 = N - \frac{2(\mu+\nu-1)}{\widehat{Z}}n_3 = N - \zeta n_3 \tag{16}$$

Interaction parameters $g$, $\nu_{12}$, $\nu_{13}$ and $\nu_{23}$, have physical representations and have to be determined by fitting Eq. (14) to experimental thermodynamic data. The equilibrium condition of $n_3$ at a specified temperature is given in Eq. (17).

$$\left(\frac{\partial G_M}{\partial n_3}\right)_{T,P,N,c_1\prime} = 0 \tag{17}$$

When Eq. (14) is substituted for $G_M$ in Eq. (17), it gives Eq. (18) (where Q is defined in Eq. (19)).

$$n_1^\mu n_2^\nu = (\mu+\nu)\, n_3\, e^{(Q-g/RT)}\, \aleph^{\mu+\nu-1} \tag{18}$$

$$Q = \frac{1}{\aleph R T}\left(-(\nu\, n_1 + \mu\, n_2)\nu_{12} + (n_1 - \mu\, n_3)\nu_{13} + (n_2 - \nu\, n_3)\nu_{23}\right.$$
$$\left.+ \frac{\zeta}{\aleph}(n_1 n_2 \nu_{12} + n_1 n_3 \nu_{13} + n_2 n_3 \nu_{23})\right) - \frac{1}{2}\widehat{Z}(\mu + \nu - \zeta)\ln\frac{\mu + \nu}{\mu + \nu - \zeta} \quad (19)$$

The thermodynamic activity of alloy components can be deduced from thermodynamic function in Eq. (20), where $i$ refers to components *A* or *B* [1,29]. Simplification of Eq. (20) for components *A* and *B* while considering the composition-dependence of $\widehat{Z}$ gives Eqs. (21) and (22). Where $Z_A$ and $Z_B$, as defined in Eq. (1), are the coordination numbers of pure liquid metals *A* and *B* at the specified *T*, respectively.

$$\ln a_i = \frac{1}{R T}\left(\frac{\partial G_M}{\partial N_i}\right)_{T,P,N} \quad (20)$$

$$\ln a_A = \ln\left(\frac{n_1}{N}\right) - \frac{1}{2}\widehat{Z}\ln\left(\frac{\aleph}{N}\right) + \frac{n_2 \nu_{12} + n_3 \nu_{13}}{\aleph R T} - \frac{n_1 n_2 \nu_{12} + n_1 n_3 \nu_{13} + n_2 n_3 \nu_{23}}{\aleph^2 R T}$$
$$- \frac{Z_A - \widehat{Z}}{2N}\left[n_3(\mu + \nu)\ln\frac{\mu + \nu}{\mu + \nu - \zeta} + N\ln\left(\frac{\aleph}{N}\right)\right] \quad (21)$$
$$- \frac{n_3 \zeta}{\aleph^2 R T}\frac{Z_A - \widehat{Z}}{\widehat{Z} N}(n_1 n_2 \nu_{12} + n_1 n_3 \nu_{13} + n_2 n_3 \nu_{23})$$

$$\ln a_B = \ln\left(\frac{n_2}{N}\right) - \frac{1}{2}\widehat{Z}\ln\left(\frac{\aleph}{N}\right) + \frac{n_1 \nu_{12} + n_3 \nu_{23}}{\aleph R T} - \frac{n_1 n_2 \nu_{12} + n_1 n_3 \nu_{13} + n_2 n_3 \nu_{23}}{\aleph^2 R T}$$
$$- \frac{Z_B - \widehat{Z}}{2N}\left[n_3(\mu + \nu)\ln\frac{\mu + \nu}{\mu + \nu - \zeta} + N\ln\left(\frac{\aleph}{N}\right)\right] \quad (22)$$
$$- \frac{n_3 \zeta}{\aleph^2 R T}\frac{Z_B - \widehat{Z}}{\widehat{Z} N}(n_1 n_2 \nu_{12} + n_1 n_3 \nu_{13} + n_2 n_3 \nu_{23})$$

Eq. (18) is solved numerically to obtain $n_3$ at different compositions of study. The corresponding $n_1$ and $n_2$ for each $n_3$ are obtained from Eqs. (7) and (8). Thereafter, $n_1$, $n_2$ and $n_3$ are used in Eqs. (14), (21) and (22) to compute $G_M$ and $a_i$s. The computed $G_M$ and $a_i$s are then compared with their experimental values, then the interaction parameters ($g, \nu_{12}, \nu_{13}, \nu_{23}$) are fitted to give satisfactory agreement between experimental and computed data. Finally, the fitted interaction parameters are used to compute other thermodynamic quantities.

Mean absolute percentage error ($Er_i$) estimation, as defined in Eq. (23), was used to measure the agreements between two sets of data [30].

$$Er_i = \pm \frac{100}{t} \sum_{i=1}^{t} \left[ \frac{d_{i,ex} - d_{i,pr}}{d_{i,ex}} \right] \qquad (23)$$

Where, $d_{i,ex}$ and $d_{i,pr}$ are the existing and predicted values, respectively, while $t$ is the size of data set.

The assumed complexes ($Al_2Sn_3$, $Al_1Zn_1$, $Sn_3Zn_2$ and $Sn_1Zn_1$), described in Table 2, were deduced from stoichiometric information [21–23]. The interaction parameters obtained for each of the binary systems and the corresponding $Er_i$ in QLT computation are also presented in Table 2. The $Er_i$ values of QLT fitting for Al-Zn at 1000 K and Sn-Zn at 750 K is lower than what was obtained in similar work on CFM [31]. The five systems can be described as weakly interacting because, their values of $g/RT$ are small compared to strongly interacting systems like Mg-Bi (16.7), Tl-Te (10.84) K-Te (47.8) and Na-Sn (22.75) [2,32]. The negative $v_{ij}s$ imply attractive interactions between the species. The plots of experimental [21–23] and computed $G_M$ and $a_i$s (Eqs. (14) and (20)), are presented in Figs. 1 and 2. The small $Er_i$ values in Table 2 and the plots presented in Figs. 1 and 2, indicate that the fitted interaction parameters are reliable. Hence, the interaction parameters, listed in Table 2, will be used in further thermodynamic studies in this work.

Table 2: Interaction parameters for each binary system and their corresponding mean absolute percentage error

| System | Exp. Ref. | $\mu$ | $\nu$ | $\dfrac{g}{RT}$ | $\dfrac{v_{12}}{RT}$ | $\dfrac{v_{13}}{RT}$ | $\dfrac{v_{23}}{RT}$ | $Er_i$ (%) $G_M$ | $a_i$s |
|---|---|---|---|---|---|---|---|---|---|
| Al-Sn at 973 K | [21] | 2 | 3 | -3.1 | 2.0 | 1.2 | -1.0 | 1.73 | 1.13 |
| Al-Zn at 973 K | [22] | 1 | 1 | -2.3 | 0.9 | 1.0 | 0.2 | 1.22 | 1.96 |
| Al-Zn at 1000 K | [21] | 1 | 1 | -2.6 | 0.9 | 1.3 | 0.2 | 1.12 | 1.72 |
| Sn-Zn at 750 K | [21] | 3 | 2 | -2.4 | 1.6 | -1.3 | 0.6 | 0.72 | 1.26 |
| Sn-Zn at 973 K | [23] | 1 | 1 | -2.5 | 0.7 | -1.0 | 1.6 | 2.42 | 3.39 |

**2.2.1 Enthalpy and entropy of formation**

The description of the enthalpy and entropy of mixing, within the quasi-lattice theory requires the knowledge of the partial differentials of the interaction parameters with respect to temperature; the experimental data at a specified temperature. From thermodynamic relations [24,25], the enthalpy of mixing, $H_M$, can be defined as given in Eq. (24). Differentiating Eq.

(14) while putting the temperature-dependent $\hat{Z}$ into consideration, and substituting in Eq. (24) gives Eq. (25), where $i, j = \{1, 2, 3\}$.

$$H_M = G_M - T \left(\frac{\partial G_M}{\partial T}\right)_P \tag{24}$$

$$H_M = -n_3 \left(g - T \frac{\partial g}{\partial T}\right) + \frac{1}{2} \frac{\partial \hat{Z}}{\partial T} R T^2 \left(n_3(\mu + \nu) \ln \frac{\mu + \nu}{\mu + \nu - \zeta} + N \ln \frac{\aleph}{N}\right) \\ + \frac{1}{\aleph} \sum_{i<j} \sum n_i n_j \left(\nu_{ij} - T \frac{\partial \nu_{ij}}{\partial T}\right) + T \frac{\partial \hat{Z}}{\partial T} \frac{n_3 \zeta}{\aleph^2 \hat{Z}} \sum_{i<j} \sum n_i n_j \nu_{ij} \tag{25}$$

The entropy of mixing, $S_M$, is defined in Eq. (26) [24,25]. A simplified expression for $S_M$, given in Eq. (27), is obtained by substituting Eqs. (14) and (25) for $G_M$ and $H_M$, respectively, in Eq. (26).

$$S_M = \frac{H_M - G_M}{T} \tag{26}$$

$$S_M = n_3 \frac{\partial g}{\partial T} - R \left(n_1 \ln \frac{n_1}{N} + n_2 \ln \frac{n_2}{N} + n_3 \ln \frac{(\mu + \nu) n_3}{N} - \frac{1}{2} \hat{Z} n_3 (\mu + \nu - \zeta) \ln \frac{\mu + \nu}{\mu + \nu - \zeta} \right. \\ \left. - \frac{1}{2} \hat{Z} \aleph \ln \frac{\aleph}{N}\right) + \frac{1}{2} \frac{\partial \hat{Z}}{\partial T} R T \left(n_3(\mu + \nu) \ln \frac{\mu + \nu}{\mu + \nu - \zeta} + N \ln \frac{\aleph}{N}\right) \\ - \frac{1}{\aleph} \sum_{i<j} \sum n_i n_j \frac{\partial \nu_{ij}}{\partial T} + \frac{\partial \hat{Z}}{\partial T} \frac{n_3 \zeta}{\aleph^2 \hat{Z}} \sum_{i<j} \sum n_i n_j \nu_{ij} \tag{27}$$

The partial differential of the average coordination number with respect to system temperature $(\partial \hat{Z}/\partial T)$ in Eqs. (24) and (27) is defined in (28).

$$\frac{\partial \hat{Z}}{\partial T} = -Z_1 c_1 \left(\frac{\Delta H_{m1}}{Z_c R T^2} + \frac{1}{T - T_{m1} + 1}\right) - Z_2 c_2 \left(\frac{\Delta H_{m2}}{Z_c R T^2} + \frac{1}{T - T_{m2} + 1}\right) \tag{28}$$

Since the interaction parameters are temperature dependent, the partial differentials $\left(\frac{\partial g}{\partial T}, \frac{\partial \nu_{12}}{\partial T}, \frac{\partial \nu_{13}}{\partial T} \text{ and } \frac{\partial \nu_{23}}{\partial T}\right)$ were fitted to conform with experimental $H_M$ and $S_M$, for Al-Sn at 973 K, Al-Zn at 1000 K and Sn-Zn at 750 K. Excess entropy of mixing $(S_M^{xs})$ was estimated from real and ideal entropies of mixing, using thermodynamic relations given in Eq. (29) [24,25].

$$S_M^{xs} = S_M + NR \sum c_i \ln c_i \tag{29}$$

### 2.2.2 Concentration-concentration fluctuations in the long-wavelength limit and Warren-Cowley short range order parameter

Concentration-concentration fluctuations in the wavelength limit ($Scc(0)$) is a microscopic function that is useful in the study of the nature of atomic order. It provides information that determines the compound forming or phase separating nature of alloys. And $Scc(0)$ can be obtained easily from thermodynamic relations defined in Eq. (30). The measured $Scc(0)$ values are compared with their ideal values ($Scc^{id}(0)$) (Eq. (31)) to make meaningful deductions. When $Scc(0) > Scc^{id}(0)$ for a certain composition, it implies a tendency for homocoordination, while $Scc(0) < Scc^{id}(0)$ implies a tendency for heterocoordination.

$$Scc(0) = RT \Big/ \left(\frac{\partial^2 G_M}{\partial c_1^2}\right)_{T,P,N} = c_2\, a_A \Big/ \left(\frac{\partial a_A}{\partial c_1}\right)_{T,P,N} = c_1\, a_B \Big/ \left(\frac{\partial a_B}{\partial c_2}\right)_{T,P,N} \tag{30}$$

$$Scc^{id}(0) = c_1 c_2 \tag{31}$$

In this work, the second-order differential in Eq. (30) was solved numerically to obtain $Scc(0)$.

The Warren-Cowley short range order parameter, $\alpha_1$, is helpful in quantifying the degree of order in the liquid alloy [33,34]. Determination of $\alpha_1$ can be made from the knowledge of the concentration-concentration and number-number structure factors from diffraction experiments. However, experimental structure factors are not easily measured during diffraction. Furthermore, $\alpha_1$ can be determined from the knowledge of $Scc(0)$ and $Scc^{id}(0)$ as defined in Eq. (35) [33–36]. When $\alpha_1 > 0$, it implies preference for like-atom pairing as nearest neighbours, whereas $\alpha_1 < 0$ corresponds to preference for unlike-atom pairing as nearest neighbours, while $\alpha_1 = 0$ denotes random distribution of atoms. The limiting values for $\alpha_1$ when $c_1 \leq \frac{1}{2}$ and $c_1 \geq \frac{1}{2}$ are given in Eqs. (32) and (33), respectively. Eqs. (32) and (33) reduces to Eq. (34) when $c_1 = c_2 = \frac{1}{2}$ [35].

$$-\frac{c_1}{c_2} \leq \alpha_1 \leq 1 \tag{32}$$

$$-\frac{c_2}{c_1} \leq \alpha_1 \leq 1 \tag{33}$$

$$-1 \leq \alpha_1 \leq +1 \tag{34}$$

When the value of $\alpha_1$ is maximum (+1), it implies complete phase separation of components in the mixture, while its minimum value (-1) implies complete ordering of unlike-atoms as nearest neighbours. The relationship between $\alpha_1$, $Scc(0)$ and $Scc^{id}(0)$ are provided in Eq. (35) [33–36].

$$\alpha_1 = \frac{\left(Scc(0)/Scc^{id}(0)\right) - 1}{Scc(0)/Scc^{id}(0)\left(\hat{Z} - 1\right) + 1} \tag{35}$$

The Average coordination number, $\hat{Z}$, in Eq. (35) was obtained using Eq. (3).

## 2.3 Surface concentration and surface tension

A statistical mechanical approach to the modelling of surface properties, using a concept of a layered structure near the interface is known to be useful in binary alloys [32,37]. The grand partition functions setup for the surface layer and the bulk provides a relation between surface and bulk compositions, given in Eq. (36) [32]. In Eq. (36), $\sigma$ is the surface tension of the mixture, $c_i^s$, $\sigma_i$, $\gamma_i$ and $\gamma_i^s$, are the surface concentration, surface tension, bulk activity coefficient and surface activity coefficient of component $i$ at temperature $T$, respectively, $A_0$ is mean surface area of the mixture (defined in Eq. (37)), $N_0$ is Avogadro's number and $k_B$ is Boltzmann's constant. The surface activity coefficient, $\gamma_i^s$, is defined in Eq. (38), where $\gamma_i(c_i^s)$ implies the use of $c_i^s$ in place of $c_i$ in the computation of activity coefficient. Similar to bulk properties (where $a_i = c_i \times \gamma_i$), surface activity ($a_i^s$) is a product of $c_i^s$ and $\gamma_i^s$.

Eq. (36) was solved numerically to obtain surface concentration and surface tension of each of the binary alloy systems.

$$\sigma = \sigma_1 + \frac{k_B T}{A_0}\ln\frac{c_1^s}{c_1} + \frac{k_B T}{A_0}\ln\frac{\gamma_1^s}{\gamma_1} = \sigma_2 + \frac{k_B T}{A_0}\ln\frac{c_2^s}{c_2} + \frac{k_B T}{A_0}\ln\frac{\gamma_2^s}{\gamma_2} \tag{36}$$

$$A_0 = 1.102 \, N_0^{-2/3} \left[ c_1 \, V_{m1}^{2/3} + c_2 \, V_{m2}^{2/3} \right] \quad (37)$$

$$\ln \gamma_i^S = p \ln \gamma_i(c_i^S) + q \ln \gamma_i \quad (38)$$

In Eq. (38), $p$ and $q$ are fractions known as surface coordination functions such that $p + 2q = 1$. They are fractions of the total number of nearest neighbours made by an atom within the layer in which it lies and that in the adjoining layer, respectively. For closed packed structures, $p = 1/2$ and $q = 1/4$.

The surface tension for pure components at specified temperatures, applied in the solution of Eq. (36), is given in Table 3.

Table 3: Surface tension for pure components

| Metal | $\sigma_i$ [a] $(N \, m^{-1})$ | | |
|---|---|---|---|
| $i$ | 750 K | 973 K | 1000 K |
| Al | - | 0.9000 | 0.8906 |
| Sn | 0.5380 | 0.5179 | - |
| Zn | 0.7723 | 0.7344 | 0.7298 |

[a] Obtained from [26]

## 2.4 Effect of approximation of coordination number on accuracy of computation

To study the effects of the approximation of coordination numbers on the accuracy of computation of thermodynamic properties, temperature- and composition-independent coordination numbers (Z) are used in computing thermodynamic properties while retaining the energy interaction parameters obtained from previous fittings.

Recall that, $\hat{Z}$ (Eq. (3)) depends on the temperature and composition of the system, as well as the structure of system's components. A temperature-independent coordination number implies that $dZ/dT = 0$ while a composition-independent coordination number implies $\hat{Z} = Z_A = Z_B = Z$.

Hence the expressions for $G_M$, $a_i s$, $H_M$, and $S_M$ (defined in Eqs. (14), (21), (22), (25) and (27)), containing average coordination numbers ($\hat{Z}$) as given in Eq. (3)) reduce to Eqs. (39) – (43), respectively. The known variables with asterisk ($G_M^*$, $a_A^*$, $a_B^*$, $H_M^*$, $S_M^*$, $\zeta^*$ and $\aleph^*$ in Eqs. (39) – (45)) indicates they were obtained using constant Z (temperature- and composition-independent coordination numbers).

In this study on approximate Z, $\hat{Z} = Z = \{2, 6, 10, 12, 10^5, 10^{10}\}$ is used to compute the thermodynamic properties across all compositions of Al-Sn at 973 K, Al-Zn at 1000 K and Sn-Zn at 750 K. The use of $Z = 2, 10^5$ and $10^{10}$ will help the results obtained via $\hat{Z}$ be comparable to the Conformal solution and Flory's approximations of the CFM [2] where Z is two and infinity [1], while Z=2, 6, 10 and 12 is applied to help compare with applications of QLT in previous works [1,38–41].

$$G_M^* = -n_3 g + RT \left[ n_1 \ln \frac{n_1}{N} + n_2 \ln \frac{n_2}{N} + n_3 \ln \frac{(\mu+\nu)n_3}{N} \right.$$
$$\left. - \frac{1}{2} Z n_3 (\mu + \nu - \zeta^*) \ln \frac{\mu+\nu}{\mu+\nu-\zeta^*} - \frac{1}{2} Z \aleph^* \ln \frac{\aleph^*}{N} \right] \quad (39)$$
$$+ \frac{1}{\aleph^*} (n_1 n_2 \nu_{12} + n_1 n_3 \nu_{13} + n_2 n_3 \nu_{23})$$

$$\ln a_A^* = \ln\left(\frac{n_1}{N}\right) - \frac{1}{2} Z \ln\left(\frac{\aleph^*}{N}\right) + \frac{n_2 \nu_{12} + n_3 \nu_{13}}{\aleph^* RT} - \frac{n_1 n_2 \nu_{12} + n_1 n_3 \nu_{13} + n_2 n_3 \nu_{23}}{\aleph^{*2} RT} \quad (40)$$

$$\ln a_B^* = \ln\left(\frac{n_2}{N}\right) - \frac{1}{2} Z \ln\left(\frac{\aleph^*}{N}\right) + \frac{n_1 \nu_{12} + n_3 \nu_{23}}{\aleph^* RT} - \frac{n_1 n_2 \nu_{12} + n_1 n_3 \nu_{13} + n_2 n_3 \nu_{23}}{\aleph^{*2} RT} \quad (41)$$

$$H_M^* = -n_3 \left(g - T \frac{\partial g}{\partial T}\right) + \frac{1}{\aleph^*} \sum_{i<j} \sum n_i n_j \left(\nu_{ij} - T \frac{\partial \nu_{ij}}{\partial T}\right) \quad (42)$$

$$S_M^* = n_3 \frac{\partial g}{\partial T} - R \left( n_1 \ln \frac{n_1}{N} + n_2 \ln \frac{n_2}{N} + n_3 \ln \frac{(\mu+\nu)n_3}{N} \right.$$
$$\left. - \frac{1}{2} Z n_3 (\mu + \nu - \zeta^*) \ln \frac{\mu+\nu}{\mu+\nu-\zeta} - \frac{1}{2} Z \aleph^* \ln \frac{\aleph^*}{N} \right) - \frac{1}{\aleph^*} \sum_{i<j} \sum n_i n_j \frac{\partial \nu_{ij}}{\partial T} \quad (43)$$

$$\zeta^* = \frac{2(\mu + \nu - 1)}{Z} \quad (44)$$

$$\aleph^* = n_1 + n_2 + \left(\mu + \nu - \frac{2(\mu+\nu-1)}{Z}\right) n_3 = N - \frac{2(\mu+\nu-1)}{Z} n_3 = N - \zeta^* n_3 \quad (45)$$

# 3 Results and discussion

Based on the theoretical formalism described in Section 2, it is important to note that the fitted interaction parameters will remain unchanged in subsequent calculations.

## 3.1 Enthalpy and entropy of formation

The partial differentials of interaction parameters were fitted for Al-Sn at 973 K, Al-Zn at 1000 K and Sn-Zn at 750 K while retaining the interaction parameters presented in Table 2. Satisfactory fits were obtained using the values presented in Table 4. The experimental and computed $H_M/RT$ and $S_M^{xs}/R$ were estimated and the results are presented as plots in Fig. 3 (a – c). The $Er_i$ values, presented in Table 4, and the $H_M/RT$ and $S_M^{xs}/R$ plots in Fig. 3 show good agreements between the fitted theoretical values and experiments.

Table 4: Partial differentials of interaction parameters obtained from experimental data [21]

| System | $\frac{\partial g}{\partial T}$ | $\frac{\partial v_{12}}{\partial T}$ | $\frac{\partial v_{13}}{\partial T}$ | $\frac{\partial v_{23}}{\partial T}$ | $Er_i$ (%) $H_M$ | $Er_i$ (%) $S_M$ |
|---|---|---|---|---|---|---|
| Al-Sn at 973 K | -1.1 R | -1.1 R | 0.5 R | -0.4 R | 4.35 | 2.94 |
| Al-Zn at 1000 K | 0.8 R | 0.0 | -0.9 R | 0.3 R | 1.62 | 0.79 |
| Sn-Zn at 750 K | -5.1 R | -1.4 R | -5.6 R | -8.2 R | 3.52 | 2.10 |

The $H_M/RT$ plots in Fig. 3 show that Al-Zn at 973 K is almost symmetric about the equiatomic composition while Al-Sn at 973 K and Sn-Zn at 750 K is slightly asymmetric in nature. Fig. 3 shows that the $H_M$ for the three systems (Al-Sn at 973 K, Al-Zn at 1000 K and Sn-Zn at 973 K) exhibit positive deviation from Raoultian behaviour across all concentration range. Also, the $S_M^{xs}$ plots in Fig. 3 are positive and asymmetric around the equiatomic composition, for the three systems considered.

## 3.2 Concentration-concentration fluctuations in the long-wavelength limit and Warren-Cowley short range order parameter

The $Scc(0)$ results for Al-Sn at 973 K, Al-Zn at 973 and 1000 K, Sn-Zn at 750 and 973 K and $Scc^{id}(0)$, computed using Eq. (30), are presented as plots in Fig. 4. Fig. 4 shows homocoordination tendency in the entire composition range of Al-Sn at 973 K, Al-Zn at 973 K, and Al-Zn at 1000 K. Fig. 4 reveals ideal behaviour in Sn-Zn at 750 K around $c_{Sn} \approx 0.8$ and heterocoordination tendency at 973 K within $0.83 \leq c_{Sn} \leq 1.00$ composition range, while other composition ranges portray homocoordination tendencies.

The $\alpha_1$ results for all the systems under study, computed using Eq. (35), are presented in Fig. 5. The $\alpha_1$ results corroborated the $Scc(0)$ results at all compositions in all the systems studied. The $\alpha_1$ plots in Fig. 5 suggests that Al-Sn at 973 K possesses a slight preference for like-atoms as nearest neighbours in the entire composition range. Al-Zn at 973 and 1000 K show similar features and trends in Fig. 5, which may be attributed to the closeness of the temperatures. Both systems have a small preference for like-atoms as nearest neighbours in their entire composition range. The preference for homocoordination is observed to increase with temperature rise. Sn-Zn exhibits random atomic order at 750 K and slight preference for heterocoordination at 973 K at similar compositions as specified by $Scc(0)$. The preference for like-atoms as nearest neighbours reduces with temperature increase except at $0.68 \leq c_{Sn} \leq 0.82$.

### 3.3 Surface concentration and surface tension

The results of surface concentrations of alloy components as functions of bulk concentrations, calculated from Eq. (36), are presented as plots in Fig. 6, which showed an expected trend, particularly, the surface concentration increases with an increase in bulk concentration for all components in all the five systems studied. From Fig. 6, it can be observed that Al-atoms segregate away from the surface in Al-Sn at 973 K, Al-Zn at 973 K and Al-Zn at 1000 K. The deficiency of Al-atoms at the surface is more pronounced in the Al-Sn system when compared to Al-Zn systems. The surface deficiency of Al-atoms in Al-Sn is very pronounced in most composition range, except $c_{Al} > 0.9$ where a sharp increase in $c_{Al}^s$ is observed. The $c_{Al}^s$ in Al-Sn agrees with the thermodynamic activity plots in Fig. 2 (a), where the positive deviation of $a_{Sn}$ from Raoults law is relatively high around $c_{Al} \approx 0.8$ ($c_{Sn} \approx 0.2$). In the Al-Zn systems, the level of surface atomic-segregation reduces with increase in temperature, in other words, the Al-Zn systems approach ideality with an increase in temperature. For the Sn-Zn systems, Fig. 6 shows segregation of Sn atoms to the surface, and the segregation of the Sn atoms is greater at 750 K than at 1000 K. The $c_{Sn}^s$ agrees with the $a_{Sn}$ plots in Fig. 2 (d and e), where the positive deviation from Raoult's law is greater at 750 K than at 1000 K.

The calculated composition dependence on surface tension for the five systems is presented as plots in Fig. 7. The $\sigma$ plots for the five systems have concave shape, and it corroborates the $c_i^S$ plots in Fig. 6 – where components with lower surface tension have higher surface enrichment. The $\sigma$ for each of the binary alloy systems ranged from that of the component with lower $\sigma_i$ to that of higher $\sigma_i$. Under ideal conditions, $\sigma$ is expected to increase (or decrease) linearly with change in concentration. The $\sigma$ for Al-Sn at 973 K deviated negatively from ideality, and the deviation is greatest at $c_{Al} \sim 0.8$. The plots show that $\sigma$ for Al-Zn and Sn-Zn reduces with increase in temperature. The $\sigma$ for Sn-Zn system appeared to be less dependent on temperature within $0.1 < c_{Sn} < 0.2$.

Due to the lack of known experimental data, the surface concentrations and surface tension calculations could not be subjected to comparison.

## 3.4 Effect of approximation of coordination number on accuracy of computation

The fitted interaction parameters were applied in Eqs. (39) – (45) while setting $Z$ as $2, 6, 10, 12, 10^5, 10^{10}$. Table 5 shows the resulting $Er_i$ in $G_M$, $a_i s$, $H_M$, and $S_M$ predictions for Al-Sn at 973 K, Al-Zn at 1000 K and Sn-Zn at 750 K at different $Z$.

Table 5 shows that the $Er_i$ values for $Z = 10^5$ and $10^{10}$ are the same for each of the systems investigated. This suggests that the predictions made with $Z = 10^5$ and $Z = 10^{10}$ are identical. Furthermore, $Z = \infty$ (which reduces QLT to Flory's approximation of CFM) can be reliably modelled with $Z = 10^5$.

Table 5 also reveals that $Er_i$ values for data predicted with $\hat{Z}$ are very close to those predicted with $Z = 10$, followed by $Z = 12, 6, \infty$ and 2. This validates the assumption by many researchers [4–8] that $Z = 10$ is a very good approximation for $\hat{Z}$. It also corroborates the results by other researchers [1,42] that the Flory's approximation of the CFM is more reasonable than its Conformal solution approximation.

Table 5: Mean absolute percentage errors in computation obtained from applying different coordination numbers

| System | $Z$* | $Er_i$ (%) | | | |
|---|---|---|---|---|---|
| | | $G_M$ | $a_i s$ | $H_M$ | $S_M$ |
| Al-Sn at 973 K | $\hat{Z}$ (Eq. (3)) | 1.73 | 1.13 | 4.35 | 2.94 |
| | 2 | 28.65 | 17.52 | 35.38 | 7.17 |
| | 6 | 5.36 | 2.03 | 5.28 | 2.82 |
| | 10 | 1.09 | 1.27 | 4.41 | 3.00 |
| | 12 | 1.58 | 1.80 | 4.42 | 3.04 |
| | 1E+05 | 5.33 | 4.42 | 5.70 | 3.40 |
| | 1E+10 | 5.33 | 4.42 | 5.70 | 3.40 |
| Al-Zn at 1000 K | $\hat{Z}$ | 1.12 | 1.72 | 1.62 | 0.79 |
| | 2 | 1.93 | 2.34 | 7.87 | 3.58 |
| | 6 | 1.19 | 1.78 | 1.78 | 1.05 |
| | 10 | 1.12 | 1.70 | 1.72 | 0.76 |
| | 12 | 1.13 | 1.69 | 1.78 | 0.82 |
| | 1E+05 | 1.15 | 1.67 | 2.34 | 1.32 |
| | 1E+10 | 1.15 | 1.67 | 2.34 | 1.32 |
| Sn-Zn at 750 K | $\hat{Z}$ | 0.72 | 1.26 | 3.52 | 2.10 |
| | 2 | 20.90 | 15.95 | 37.00 | 6.58 |
| | 6 | 3.31 | 1.75 | 5.89 | 2.30 |
| | 10 | 0.77 | 1.35 | 3.49 | 1.99 |
| | 12 | 1.30 | 1.70 | 3.86 | 1.99 |
| | 1E+05 | 4.15 | 3.99 | 9.13 | 2.53 |
| | 1E+10 | 4.15 | 3.99 | 9.13 | 2.53 |

\* $\hat{Z}$ is average coordination number (composition- and temperature-dependent) defined by Eq. (3)

## 4 Conclusions

The QLT calculations have been modified by considering the dependence of coordination numbers of materials on component-structure, composition and temperature, which is based on the concept of coordination numbers [9–16]. The modified QLT has been successfully applied to study the bulk and surface properties of Al-Sn, Al-Zn and Sn-Zn binary alloys. Moreover, this work has also shown that 10 is a good approximate value for coordination number in liquid alloys.

The binary alloys were studied at temperatures corresponding to popular experimental thermodynamic database [21], and at 973 K which corresponds to the available experimental information on Al-Sn-Zn [17]. EMF data [22,23] and Gibbs-Duhem integration [24,25] were used to obtain experimental information on Al-Zn and Sn-Zn at 973 K. After satisfactory theoretical fitting for $G_M$ and $a_i s$, the modified QLT was used to theoretically fit $H_M$ and $S_M$. Thereafter, the $Scc(0)$, $\alpha_1$, $c_i^s$ and $\sigma$ for five systems (Al-Sn at 973 K, Al-Zn at 973 and 1000 K, Sn-Zn at 750 and 973 K) were deduced.

The $H_M$ for the Al-Sn, Al-Zn and Sn-Zn systems exhibit positive deviation from Raoultian behaviour while Al-Zn appear symmetric about the equiatomic composition. The $S_M^{xs}$ for the three systems appear asymmetric about the equiatomic composition. The binary alloys show homocoordination tendencies, except in Sn-Zn where ideal behaviours and slight heterocoordination tendencies is observed in the Sn-rich regions at 750 and 973 K, respectively. The surfaces of Al-Sn and Al-Zn systems are deficient of Al-atoms, while the surfaces of Sn-Zn are enriched with Sn-atoms. Components with the lower surface tension have the highest surface enrichment, in each of the binary systems. The $Scc(0)$ and $\alpha_1$ results for Al-Sn at 973 K, Al-Zn at 1000 K and Sn-Zn result at 750 K agree with ref [18].

This work has improved the knowledge on the binary subsystems of Al-Sn-Zn (a lead-free solder), thereby contributing to the thermodynamic database of lead-free solders. It has also laid a foundation for the QLT studies on Al-Sn-Zn and other ternary systems.

# Acknowledgement

OM is grateful to Prof. Adam Dębski of the Institute of Metallurgy and Materials Science, Polish Academy of Sciences, Krakow, Poland, for his assistance with relevant literature.

# References

[1]  A.B. Bhatia, R.N. Singh, A quasi-lattice theory for compound forming molten alloys, Phys. Chem. Liq. 13 (1984) 177–190. https://doi.org/10.1080/00319108408080778.

[2]  A.B. Bhatia, W.H. Hargrove, Concentration fluctuations and thermodynamic properties of some compound forming binary molten systems, Phys. Rev. B. 10 (1974) 3186. https://doi.org/10.1103/PhysRevB.10.3186.

[3]  R. Novakovic, D. Giuranno, E. Ricci, S. Delsante, D. Li, G. Borzone, Bulk and surface properties of liquid Sb-Sn alloys, Surf. Sci. 605 (2011) 248–255. https://doi.org/10.1016/j.susc.2010.10.026.

[4]  D. Adhikari, R.P. Koirala, B.P. Singh, Thermodynamic , Structural and Transport Properties of Molten Copper-Thallium Alloys, Int. J. Math. Comput. Phys. Electr. Comput. Eng. 7 (2013) 792–797.

[5]  S.K. Yadav, L.N. Jha, I.S. Jha, B.P. Singh, R.P. Koirala, D. Adhikari, Prediction of thermodynamic and surface properties of Pb−Hg liquid alloys at different temperatures, Philos. Mag. 96 (2016) 1909–1925. https://doi.org/10.1080/14786435.2016.1181281.

[6]  A. Kumar, I.S. Jha, B.P. Singh, Thermodynamics and atomic order in molten Mg-Bi alloy, Adv. Mater. Lett. 4 (2013) 155–159. https://doi.org/10.5185/amlett.2013.8541.

[7]  A. Dębski, M. Zabrocki, W. Gąsior, Calorimetric study and thermodynamic description of liquid In-Li alloys, J. Mol. Liq. 243 (2017) 72–77. https://doi.org/10.1016/j.molliq.2017.08.022.

[8]  D. Adhikari, I.S. Jha, B.P. Singh, Transport and surface properties of molten Al-Mn alloy, Adv. Mater. Lett. 3 (2012) 226–230. https://doi.org/10.5185/amlett.2012.3324.

[9]  S.H. Simon, The Oxford solid state basics, Oxford University Press, Oxford, 2013.

[10]  P. Hofmann, Solid State Physics: An Introduction, 2nd ed., Wiley-VCH, Weinheim, 2015.

[11] N.S. Gingrich, L. Heaton, Structure of Alkali Metals in the Liquid State, J. Chem. Phys. 34 (1961) 873–878. https://doi.org/10.1063/1.1731688.

[12] J.R. Wilson, The structure of liquid metals and alloys, Metall. Rev. 10 (1965) 381–590. https://doi.org/10.1179/mtlr.1965.10.1.381.

[13] J.R. Cahoon, The first coordination number for liquid metals, Can. J. Phys. 82 (2004) 291–301. https://doi.org/10.1139/p04-003.

[14] D.P. Tao, Prediction of the coordination numbers of liquid metals, Metall. Mater. Trans. A. 36 (2005) 3495–3497. https://doi.org/10.1007/s11661-005-0023-5.

[15] Y. Waseda, The structure of non-crystalline materials: Liquids and amorphous solids, McGraw-Hill, New York, 1980.

[16] G. Saffarini, Glass transition temperature and molar volume versus average coordination number in Ge100-xSx bulk glasses, Appl. Phys. A Solids Surfaces. 59 (1994) 385–388. https://doi.org/10.1007/BF00331716.

[17] S. Knott, A. Mikula, Thermodynamic properties of liquid Al-Sn-Zn alloys: A possible new lead-free solder material, Mater. Trans. 43 (2002) 1868–1872. https://doi.org/10.2320/matertrans.43.1868.

[18] L.C. Prasad, A. Mikula, Thermodynamics of liquid Al-Sn-Zn alloys and concerned binaries in the light of soldering characteristics, Phys. B Condens. Matter. 373 (2006) 64–71. https://doi.org/10.1016/j.physb.2005.11.073.

[19] D.P. Tao, Prediction of activities of all components in the lead-free solder systems Bi-In-Sn and Bi-In-Sn-Zn, J. Alloys Compd. 457 (2008) 124–130. https://doi.org/10.1016/j.jallcom.2007.02.123.

[20] Y.A. Odusote, A.I. Popoola, K.D. Adedayo, S.T. Ogunjo, Thermodynamic properties of Al in ternary lead-free solder Al-Sn-Zn alloys, Mater. Sci. 35 (2017) 583–593. https://doi.org/10.1515/msp-2017-0080.

[21] R. Hultgren, P.D. Desai, D.T. Hawkins, M. Geiser, K.K. Kelley, eds., Selected values of the thermodynamic properties of binary alloys, ASM, Metals Park, OH., 1973.

[22] B. Predel, U. Schallner, Beitrag zur Kenntnis der thermodynamischen Eigenschaften binarer flussiger Legierungen des Aluminiums mit Gallium und Zink, Zeitschrift Für Met. 60 (1969) 869.

[23] Z. Moser, W. Gąsior, Thermodynamic studies of Zn-Sn liquid solutions, Bull. Polish Acad. Sci. 31 (1983) 19–25.

[24] J.-P. Ansermet, S.D. Brechet, Principles of Thermodynamics, Cambridge University Press, Cambridge, 2019. https://doi.org/10.1017/9781108620932.

[25] R.F. Sekerka, Thermal Physics: Thermodynamics and Statistical Mechanics for Scientists and Engineers, Elsevier, Waltham, 2015. https://doi.org/10.1016/C2014-0-03233-9.

[26] T. Iida, R.I.L. Guthrie, The physical properties of liquid metals, Clarendon Press, Oxford, 1988.

[27] G. Saffarini, A. Schlieper, Average coordination-number dependence of glass-transition temperature in Ge-In-Se chalcogenide glasses, Appl. Phys. A Mater. Sci. Process. 61 (1995) 29–32. https://doi.org/10.1007/BF01538206.

[28] G. Saffarini, A. Saiter, M.R. Garda, J.M. Saiter, Mean-coordination number dependence of the fragility in Ge-Se-In glass-forming liquids, Phys. B Condens. Matter. 389 (2007) 275–280. https://doi.org/10.1016/j.physb.2006.06.163.

[29] R. Novakovic, T. Tanaka, Bulk and surface properties of Al-Co and Co-Ni liquid alloys, Phys. B Condens. Matter. 371 (2006) 223–231. https://doi.org/10.1016/j.physb.2005.10.111.

[30] U. Khair, H. Fahmi, S. Al Hakim, R. Rahim, Forecasting Error Calculation with Mean Absolute Deviation and Mean Absolute Percentage Error, in: J. Phys. Conf. Ser., Institute of Physics Publishing, 2017. https://doi.org/10.1088/1742-6596/930/1/012002.

[31] O.E. Awe, O.M. Oshakuade, Computation of infinite dilute activity coefficients of binary liquid alloys using complex formation model, Phys. B Condens. Matter. 487 (2016) 13–17. https://doi.org/10.1016/j.physb.2016.01.023.

[32] O. Akinlade, R.N. Singh, Bulk and surface properties of liquid In–Cu alloys, J. Alloys Compd. 333 (2002) 84–90. https://doi.org/10.1016/S0925-8388(01)01733-9.

[33] B.E. Warren, X-Ray Diffraction, Addison-Wesley, Reading, 1969.

[34] J.M. Cowley, An approximate theory of order in alloys, Phys. Rev. 77 (1950) 669–675. https://doi.org/10.1103/PhysRev.77.669.

[35] R.N. Singh, Short-range order and concentration fluctuations in binary molten alloys, Can. J. Phys. 65 (1987) 309–325. https://doi.org/10.1139/p87-038.

[36] R.N. Singh, D.K. Pandey, S. Sinha, N.R. Mitra, P.L. Srivastava, Thermodynamic properties of molten LiMg alloy, Phys. B+C. 145 (1987) 358–364. https://doi.org/10.1016/0378-4363(87)90105-7.

[37] O. Akinlade, O.E. Awe, Bulk and surface properties of liquid Ga-Tl and Zn-Cd alloys, Int. J. Mater. Res. 97 (2006) 377–381. https://doi.org/10.3139/146.101227.

[38] O. Akinlade, F. Sommer, Concentration fluctuations and thermodynamic properties of ternary liquid alloys, J. Alloys Compd. 316 (2001) 226–235. https://doi.org/10.1016/S0925-8388(00)01450-X.

[39] O.E. Awe, O. Akinlade, L.A. Hussain, A Quasi-Lattice Theory for Compound Forming Ternary Liquid Alloys, Int. J. Mod. Phys. B. 20 (2006) 3319–3340. https://doi.org/10.1142/S0217979206035412.

[40] R. Novakovic, E. Ricci, F. Gnecco, D. Giuranno, G. Borzone, Surface and transport properties of Au-Sn liquid alloys, Surf. Sci. 599 (2005) 230–247. https://doi.org/10.1016/j.susc.2005.10.009.

[41] E.A. Guggenheim, Mixtures, Oxford University Press, London, 1952.

[42] O.E. Awe, O. Akinlade, L.A. Hussain, A quasi-lattice theory for compound forming ternary liquid alloys, Int. J. Mod. Phys. B. 20 (2006) 3319–3340. https://doi.org/10.1142/S0217979206035412.

# Figures

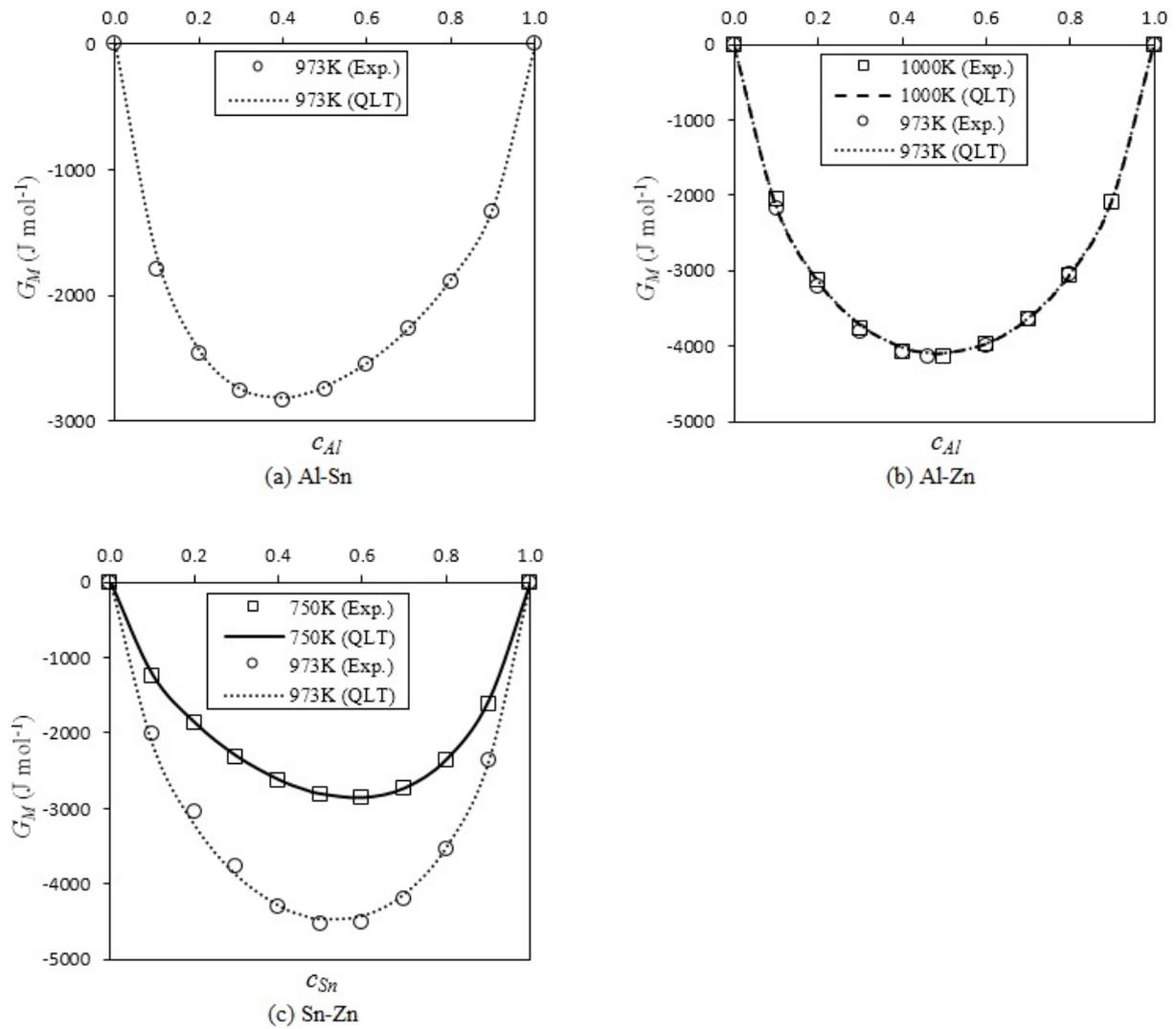

Fig. 1: $G_M$ in Al-Sn at 973 K, Al-Zn at 973 and 1000 K, and Sn-Zn at 750 and 973 K

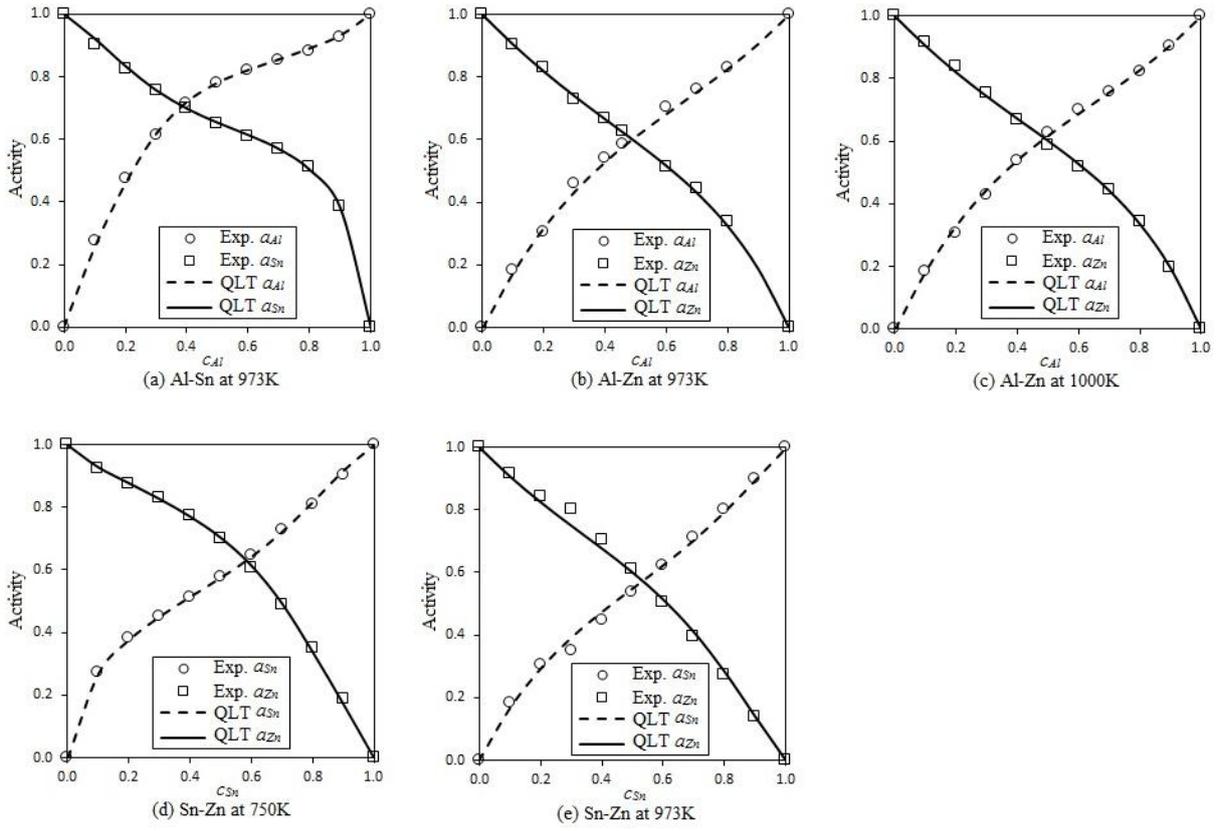

Fig. 2: Activities of all components in Al-Sn at 973 K, Al-Zn at 973 and 1000 K, and Sn-Zn at 750 and 973 K

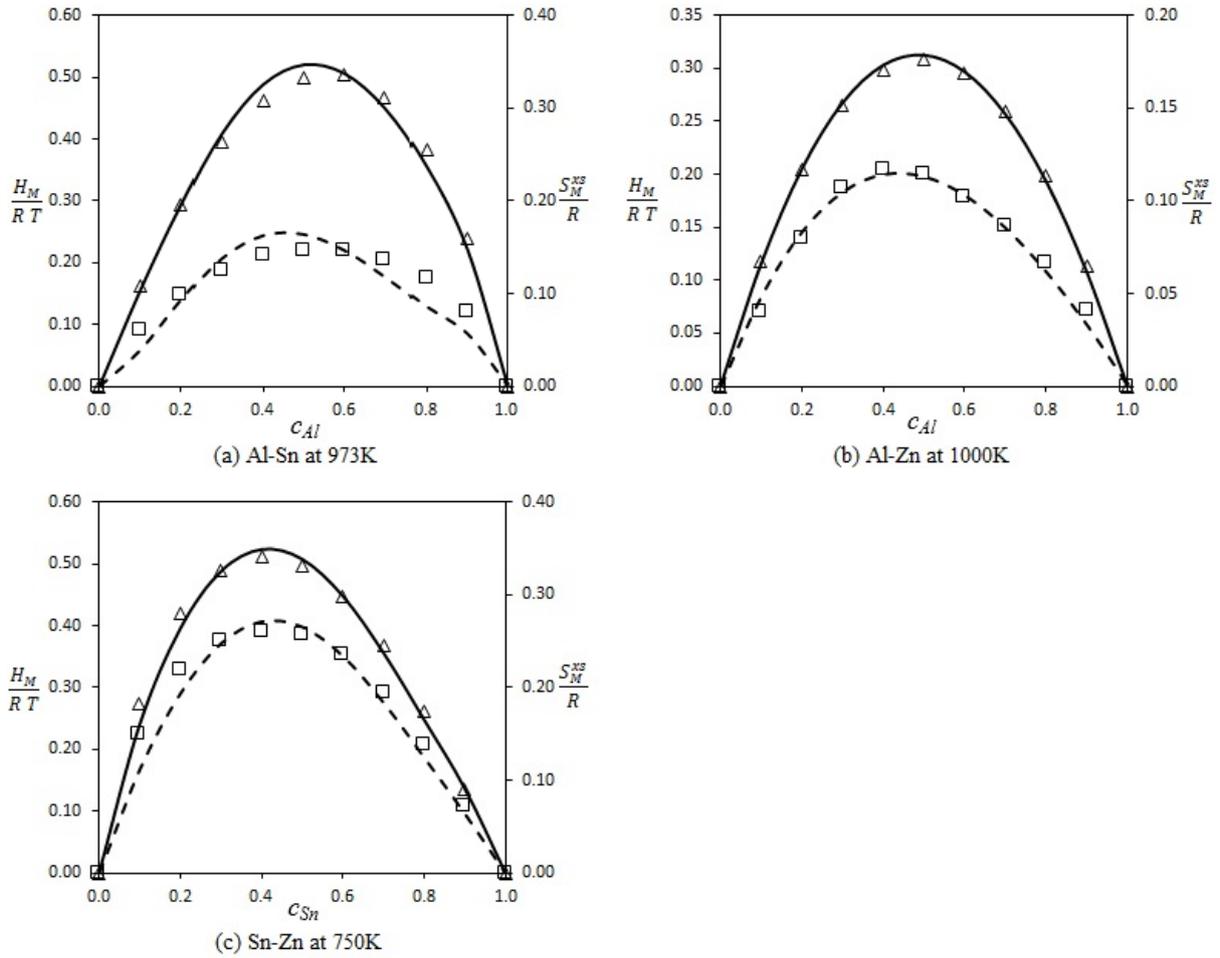

Fig. 3: Concentration dependence of $H_M/RT$ and $S_M^{xs}/R$. Triangles represent $H_M/RT$ obtained from experiments [21], squares represent $S_M^{xs}/R$ obtained from experiments [21], solid lines represent QLT-computed $H_M/RT$ while dashed lines represent QLT-computed $S_M^{xs}/R$

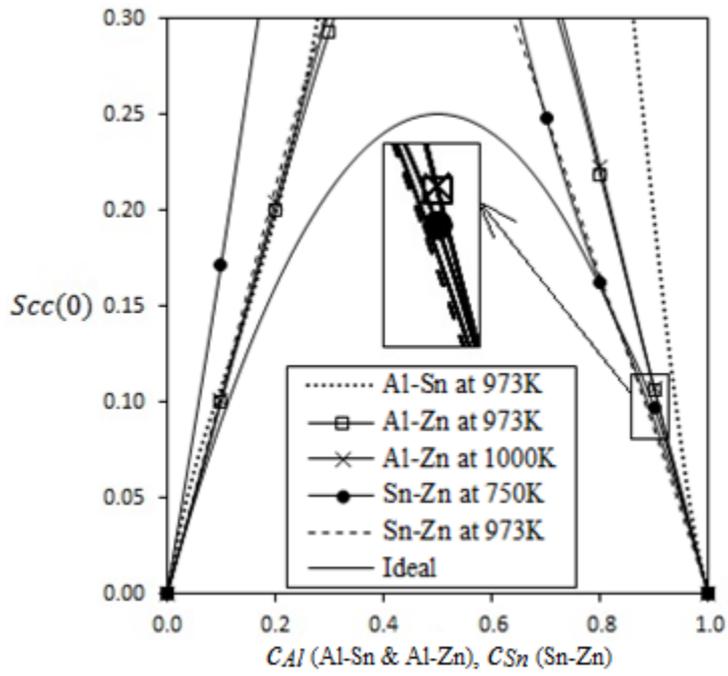

Fig. 4: $Scc(0)$ for Al-Sn at 973 K, Al-Zn at 973 and 1000 K, and Sn-Zn at 750 and 973 K

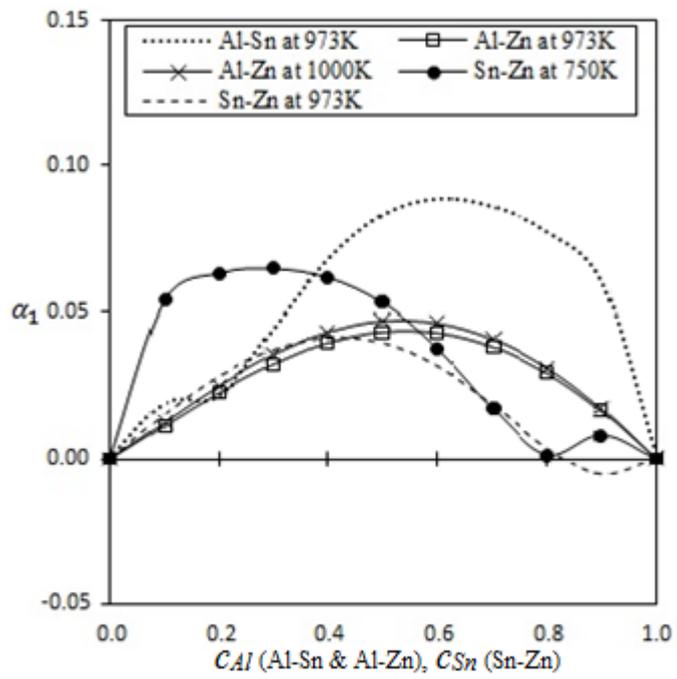

Fig. 5: $\alpha_1$ for Al-Sn at 973 K, Al-Zn at 973 and 1000 K, and Sn-Zn at 750 and 973 K

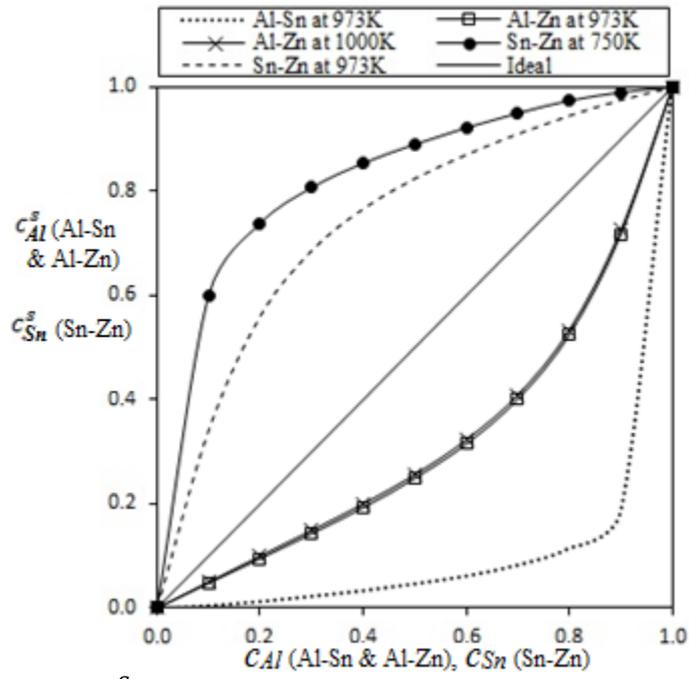

Fig. 6: $c_{Al}^S$ vs. $c_{Al}$ for Al-Sn at 973 K, Al-Zn at 973 and 1000 K, and $c_{Sn}^S$ vs. $c_{Sn}$ for Sn-Zn at 750 and 973 K

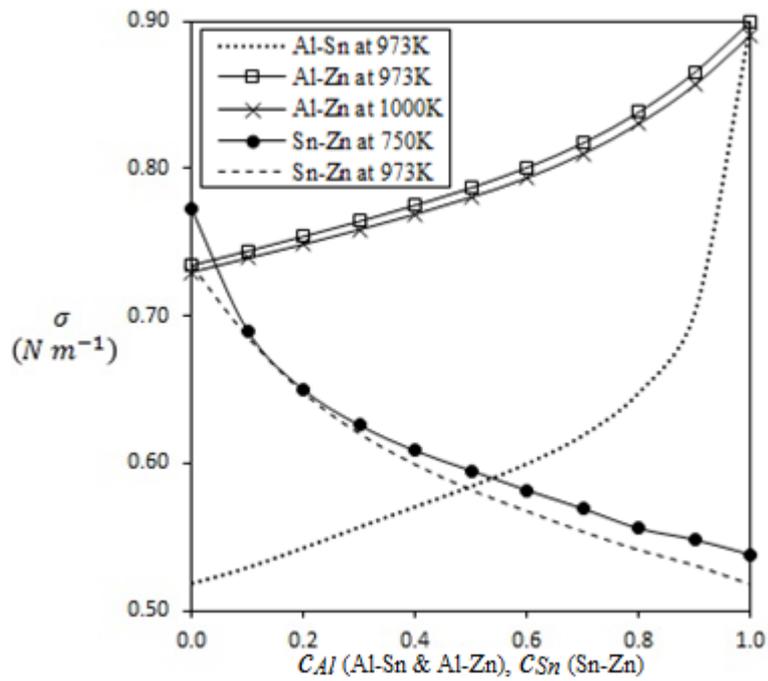

Fig. 7: Concentration dependence of surface tension for Al-Sn at 973 K, Al-Zn at 973 and 1000 K, and Sn-Zn at 750 and 973 K